# The Distant Possibility of Using a High-Luminosity Muon Source to Measure the Mass of the Neutrino Independent of Flavor Oscillations

By John Michael Williams

`jwill@AstraGate.net`
Markanix Co.
P. O. Box 2697
Redwood City, CA 94064

2001 February 19 (*v.* 1.02)

**Abstract**: Short-baseline calculations reveal that if the neutrino were massive, it would show a beautifully structured spectrum in the energy difference between storage ring and detector; however, this spectrum seems beyond current experimental reach. An interval-timing paradigm would not seem feasible in a short-baseline experiment; however, interval timing on an Earth-Moon long baseline experiment might be able to improve current upper limits on the neutrino mass.

## Introduction

After the Kamiokande and IMB proton-decay detectors unexpectedly recorded neutrinos (probably electron antineutrinos) arriving from the 1987A supernova, a plethora of papers issued on how to use this happy event to estimate the mass of the neutrino. Many of the estimates based on these data put an upper limit on the mass of the electron neutrino of perhaps 10 $\text{eV}/c^2$ [1].

When Super-Kamiokande and other instruments confirmed the apparent deficit in electron neutrinos from the Sun, and when a deficit in atmospheric muon-neutrinos likewise was observed, this prompted the extension of the kaon-oscillation theory to neutrinos, culminating in a flavor-oscillation theory based by analogy on the CKM quark mixing matrix. The oscillation theory was sensitive enough to provide evidence of a neutrino mass, even given the low statistics available at the largest instruments.

However, there is reason to doubt that the CKM analysis validly can be applied physically over the long, nonvirtual propagation distances of neutrinos [2].

The relatively high luminosity of a muon storage ring as a neutrino factory raises the question of a possible alternative estimate of the neutrino mass, based on the kind of kinematic analysis attempted with the SN1987A supernova data. We present here some preliminary calculations describing how this estimate might be made.

Both muon and electron neutrinos are produced by such storage rings, and we do not rule out a difference in massiveness between the two types. If mass could be confirmed in either of the suggested paradigms, adjustment of the circulating-muon polarization (which would vary the electron neutrino flux [12; 13]) would allow the mass difference to be estimated.

The nonoscillating approaches discussed here require noncontinuous neutrino production, corresponding to a storage ring with a circulating train of muons at any instant filling only part of the ring, or to a storage ring with shuntable barrel [3].

# Problem Overview

## *Geometry Considerations*

We first sketch a generalization of the muon storage ring as an oval, ignoring the alternative shapes under consideration, bowtie (e. g., [3; 7]) or triangle (e. g., [4; 7]), as not bearing on the experiment we propose.

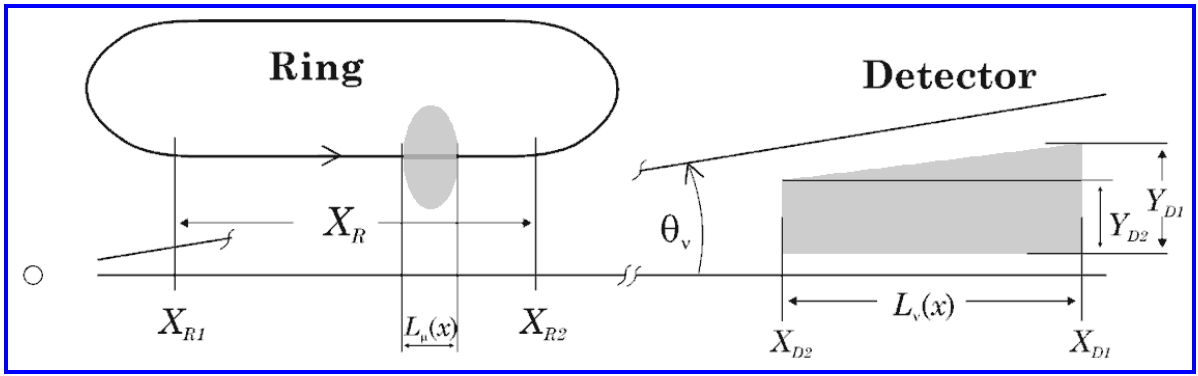

**Figure 1.  Muon ring neutrino source and detector, showing the spatial variables used for neutrino mass analysis in the text.  The direction of *x* is on the line between barrel and detector; *y* is transverse.**

As in Fig. 1, assume a storage ring of circumference $C$ and a straight run $R$ of fixed length $X_R = |X_{R2} - X_{R1}|$. We call this straight run, the <u>barrel</u> of the ring, because it is assumed aimed at a target, the detector $D$. We assume a barrel length $X_R = C/3$.

Leaving Lorentz factors for later, while a muon bunch with lab-frame length $L_m(x)$ is in the barrel, it will accumulate a bunch of neutrinos which will arrive at the detector as an image bunch of lab-frame length $L_n(x)$, as shown in Fig. 1.

Noticing the origin circle $O$ at far left in Fig. 1, we assume each muon bunch to decay and create a neutrino-bunch element of constant (transverse) radius $R_R$ in the barrel; then, using the opening half-angle $q_n$, we may project back from any location $x_R$ in the barrel, $X_{R1} \leq x_R \leq X_{R2}$, to a square-law neutrino bunch-element virtual origin $O(x_R)$ at,

$$\tan q_n = R_R/(x_R - O(x_R)) \quad \Rightarrow \quad O(x_R) = x_R - R_R/\tan q_n. \tag{1}$$

The height $Y(x_D)$, which would be the radius in cylindrical coordinates, of the detector event $x_D$, $X_{D2} \leq x_D \leq X_{D1}$, corresponding to the neutrino-bunch element event at barrel location $x_R$, would be

$$Y(x_D) = (x_D - x_R)\tan q_n. \tag{2}$$

The virtual origin $O$ in Fig. 1 will be the farther left of the bunch in question, the higher the beam energy. A muon bunch at some location $x_R$, with average radius $\langle Y(x_R) \rangle = 1$ cm and $g = 500$, will be about 5 meters to the right of its $O$. The virtual origin, of course, will traverse the barrel with its respective bunch.

## *Timing Considerations*

Now we look at some of the temporal factors. We call a muon *train* a bunch of muons or a succession of such bunches so closely spaced that neutrinos from adjacent bunches in the train can not be discriminated in a single pass at the detector. Call $L_m$ the lab-frame length of this train. Call $T_R$ the lab-frame time spent by the train while any part of it was in the barrel. We wish to optimize time for the detector to distinguish events from different trains, so we assume just one train in the storage ring.

We may write the value of $T_R$ as a function of barrel length as,

$$T_R = \frac{X_R + L_m}{\langle u_{\text{muon}} \rangle} \equiv \frac{|\{x_B\}|}{\langle u_{\text{muon}} \rangle}, \tag{3}$$

defining the light-cone interval in the barrel as the length of the set of all locations $x_B$ in the barrel, for $\langle u_{\text{muon}} \rangle$ the average lab-frame speed of the muon in the barrel. Approximating the lab-frame speed of a 50 GeV muon as *c*, and assuming a 0.5-km barrel in a 1.5-km ring, the lab-frame cycle time $T$ for a stored muon will be about 5 $ms$. The relevant ring temporal parameters are illustrated in Table 1.

**Table 1. Lab-frame times for a solitary muon train of a given lab-frame length, for a 1.5 km ring with 0.5 km barrel.  Time in the barrel includes time for any part of the train.**

| Train Length $L_m$ (m) | Time in Barrel $T_R$ ($ms$) | Time Outside Barrel $T - T_R$ ($ms$) |
|---|---|---|
| ~0 | 1.7 | 3.3 |
| 125 | 2.1 | 2.9 |
| 250 | 2.5 | 2.5 |
| 375 | 2.9 | 2.1 |
| 500 | 3.3 | 1.7 |
| 625 | 3.8 | 1.4 |
| 750 | 4.2 | 0.8 |
| 875 | 4.6 | 0.4 |
| 1000 | 5.0 | ~0 |

Looking at Table 1, to perform this experiment without shunting or interrupting the stored muons, discrimination at the detector of neutrino events separated by perhaps 1 $ms$ in the lab frame is required to verify the occurrence of a train.

In the following, we refer to lab times $t$ during which at least part of the muon train is in the barrel as "the $t \subset T_R$ regime" (regime = phase-interval) of the storage ring operation. The corresponding light-cone interval in the lab frame will be discussed later as the virtual barrel length $|\{x_B\}|$ corresponding to, but greater than, $X_R$ in Fig. 1. The image will be assumed processed at the detector so that on the light cone, each detection event at point $x_D$ is superposable on the corresponding creation event at $x_B$ by simple translation.

## *Exploratory Calculations*

We look at the magnitude of the problem before attempting an experimental protocol.

### Coarse Estimate of Muon Luminosity Decay

We calculate the effect of the decay of the muons stored in the barrel on the expected neutrino population, to account for it and evaluate it as a possible systematic error.

The Lorentz factor $g$ for a 50 GeV muon with rest mass 106 MeV$/c^2$ is given by,

$$g = E/(mc^2) = 50 \cdot 10^9 / (106 \cdot 10^6) \cong 470 ; \qquad (4)$$

so, from a muon proper-time decay constant $t_P$ of some 2.2 *ms*, we can expect a muon lab-frame lifetime $t_L = g t_P$ of about $470 \cdot 2.2 \cong 1000$ *ms* ($\cong 200$ cycles). The lab-time decline in the muon population at this energy while in the barrel therefore will be given by the fraction, $1 - \exp(-(1/3)/200) = 1 - \exp(-1/600)$, or, about 0.17%. This implies a neutrino luminosity decline of less than 0.2% during any one traversal of the barrel.

Using a linear approximation, we then expect the observed mean position $\langle x \rangle$ in the barrel of all neutrino creation events to be shifted toward $X_{R1}$ by about $250 \cdot 0.0017/2$ m, or, about 21 cm.

Over a 5 km propagation distance (below) on the light cone, 21 cm represents a factor of $g \cong 110$; at the anticipated energies, the rest mass of a particle at $g = 110$ would exceed that of a muon. This means that the effect of muon decay in the barrel on $\langle x \rangle$ is far greater than the effect of any reasonable uncertainty in the mass of the neutrino. The decay in the barrel then will require a major systematic correction if our measurement is to succeed.

### Fine Estimate of Muon Luminosity Decay

Looking at Fig. 1, assuming continuously refreshed circulating muons until reaching the barrel, the corresponding rate of good-neutrino creation events will rise linearly while the train enters the barrel, will be distributed almost uniformly on the 1/2 km domain of the barrel while the train is fully contained in the barrel, and will decline linearly again as the train exits the barrel. Thus, given a barrel $\{x_R\} \subset (X_{R1}, X_{R2})$, the intensity $I(\{x_B\})$ of these events during the $t \subset T_R$ regime will be described by three linear functions on the barrel length, $I(\{x_{\text{rise}}\})$, $I(\{x_{\text{uniform}}\})$, and $I(\{x_{\text{fall}}\})$. If we decided to choose a muon train which gave us the 50% duty cycle in Table 1, $L_m(x)$ would be 250 m; so, on the light cone, $L_n(x_D)$ will be composed of a 250-m rise phase, a 250-m uniform phase, and a 250-m fall phase. Muons will not be refreshed during any of these phases, suggesting a greater effect of decay than assumed in the coarse estimate above.

To find the change in mean because of muon decay in the barrel, we superpose phases and map them back to the barrel centroid:

$$\langle x_B \rangle = \frac{1}{|\{x_B\}|} \int_{\{x_B\}} x \cdot I(\{x\}) e^{-\frac{x}{u\tau}}$$

$$= \frac{1}{|\{x_B\}|} \left[ \int_{\{x_{\text{rise}}\}} x \cdot I(\{x\}) e^{-\frac{x}{u\tau}} + \int_{\{x_{\text{uni.}}\}} x \cdot I(\{x\}) e^{-\frac{x}{u\tau}} + \int_{\{x_{\text{fall}}\}} x \cdot I(\{x\}) e^{-\frac{x}{u\tau}} \right]; \quad (5)$$

in which the exponential represents the decay rate of the muon flux, given the proper-time decay constant $\tau = 2.2\,\mu s$ at lab-frame circulation speed *v*.

Assuming relativistic muons, we use $u\gamma = c\gamma\sqrt{1-(1/\gamma)^2} = c\sqrt{\gamma^2-1}$ for $u\gamma$ in (5) and combine the superposed terms to write,

$$\langle x_B \rangle = \frac{1}{250 \cdot 750} \left[ \int_0^{250} dx \cdot x^2 \exp\left(\frac{-x}{\tau \cdot c\sqrt{\gamma^2-1}}\right) + 250 \int_{250}^{500} dx \cdot x \exp\left(\frac{-x}{\tau \cdot c\sqrt{\gamma^2-1}}\right) \right. $$
$$\left. + 750 \int_{500}^{750} dx \cdot x \exp\left(\frac{-x}{\tau \cdot c\sqrt{\gamma^2-1}}\right) - \int_{500}^{750} dx \cdot x^2 \exp\left(\frac{-x}{\tau \cdot c\sqrt{\gamma^2-1}}\right) \right]. \quad (6)$$

For 50 GeV muons, we now find that muon decay has shifted the mean neutrino creation point to $\langle x \rangle \cong 249.64$ m in the barrel, about 36 cm from the light cone mean.

We note that our new estimate of the effect of muon decay in the 500 m barrel has made the coarse luminosity decay estimate *ca.* Eq. (4) worse. It is discouraging that an experiment depending on the mean neutrino creation point in the barrel merely might provide an uncertainty complementary to that known of the lifetime of the muon, but we shall proceed anyway.

On the bright side, mass of the neutrino should not complicate use of a muon storage ring to refine measurement of the decay time of muons.

# Mass Measurement Paradigms

There are two distinct paradigms that might be used to detect and measure the mass of the neutrino: (*a*) A spectral difference method; and, (*b*) an absolute time of flight method.

## *General Assumptions*

The two paradigms below imply that calibration will be performed in air to keep systematic errors down. This requires the distance from storage ring to detector to be under about 5 km, a reasonable line-of-sight limit. Extension of the computations below to detectors at 5,000 km distance or more, is straightforward, because the distance enters as a simple extrapolation factor in all formulae. However, in the author's opinion, the effect of matter on the propagation speed of Standard-Model neutrinos is as big a question as their mass.

Temporal resolution at the detector may be defined spatially, recalling that the neutrinos will be ultrarelativistic and that $c$ maps to about $1/3$ meter per nanoscond, or $1/3$ millimeter per picosecond.  The possibility of liquid argon time-projection chambers with wire spacing on the order of a few millimeters [8] or of dense CCD vertex reconstruction to a few microns [11] suggests that picosecond final-event resolution, and thus nanosecond time-difference resolution, might be attained fairly easily.

To estimate feasibility, we assume a detector system capable of fully reconstructing $10^7$ events per year per kg mass [8].  We assume that the energy of each neutrino, for a good detector event, will be correct within 20%: For a 50 GeV neutrino, the Heisenberg uncertainty principle then allows a time resolution down to $\Delta t = \hbar/(2\Delta E) \cong 10^{-23.5}$ s ($10^{-5.5}$ as).  We guess that the system will provide a time resolution to 1 ns.  We assume some $10^{10}$ good events per year [8; 11]; for each of these events, we assume data correlated with other detector or storage-ring events as required by the paradigm under discussion.

## *Spectral Differential Time Paradigm*

In this case, we consider a number of detector events (images of barrel events) each characterized by a known distribution in time.  Perhaps such an event would be synchronized to the arrival in the barrel of a train of stored muons.  We look at the detector sample shape of the image spectrum, hoping to see a systematic departure from that at the barrel.  If there was an increase in variance, or a change in shape, it would indicate dispersion during flight, differential speed among the individual particles, and therefore nonzero mass.

We assume the muon population in the storage ring is being refreshed on each storage-ring cycle, to replace all decayed muons.  We also assume, with little loss of generality, that the storage ring puts the barrel in use exactly half of each cycle (50% duty cycle at the detector); this corresponds to the schedule in the third row of Table 1, which means a muon train 250 m long in a 1500 m ring with 500 m barrel.

As in Fig. 1.1 of [8], we have a 50 GeV muon storage ring and a $1/g$ short-baseline detector at a distance $\Delta x = 5$ km.  So, our good neutrino events each will have a fairly symmetrical, bell-shaped energy spectrum centered a little over 25 GeV.  For the present exploratory purpose, we shall read [8], Fig. 1.1, as describing a Gaussian spectrum with mean $\langle E \rangle = 27$ GeV and standard deviation $\sqrt{\langle ^2 E \rangle} = 8$ GeV.  Whenever we evaluate this spectrum, we shall cut it off at $1 \leq E \leq 50$ GeV;

this leaves about 3 standard deviations on each side of the mean and ignores the lowest-energy neutrinos, which may escape the detector anyway.

We shall abbreviate a Gaussian function of this kind as $\mathfrak{G}(E)$, merely naming the argument. The abbreviation represents a function--possibly of a random variable if so stated. To avoid any ambiguity, we define here, that

$$\mathfrak{G}(E) \equiv \mathfrak{G}\left(E, \langle E \rangle, \langle ^2 E \rangle\right) \equiv \frac{1}{\sqrt{2p\langle ^2 E \rangle}} \exp\left(-\frac{(E - \langle E \rangle)^2}{2\langle ^2 E \rangle}\right). \tag{7}$$

In the present paradigm, instead of averaging individual neutrino times of flight, we shall look for a change of shape of the neutrino event-count spectrum $\Psi_R(x, E)$ at the barrel *vs* the corresponding image spectrum, $\Psi_D(x, E)$, in the detector. To see how this would work in terms of Fig. 1, for each storage-ring cycle, so long as a train $L_m(x_B)$ was at a point $x_R$ in the barrel, the neutrino spectrum would be cumulated in the detector by a time-correlating system which would associate good events with the expected $L_n(x_D)$, the association being the geometrical $x_R \leftrightarrow x_B$ at a phase point in the cycle; or, equivalently, $x_B \leftrightarrow x_D$ during $t \subset T_R$ on the light cone. If neutrinos were massive, the detector event locations $x_D$ all would shift toward lower values than on the light cone, the lower energy neutrinos shifting farther than those of higher energy. This is sketched in Fig. 2.

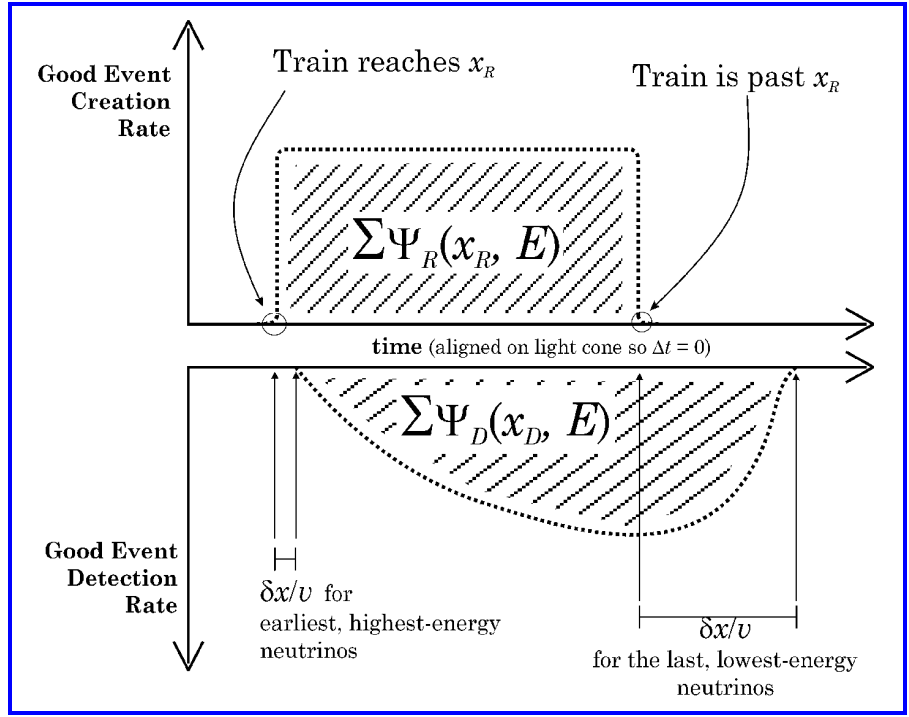

**Figure 2.  The spectral differential-time paradigm:  Sketch of barrel and image events at a fixed point $x_R$ in the storage ring barrel.  Time $t$ is in lab time, $x \equiv -ct$.  The total energy flux from barrel to image is conserved for the good-event neutrino population.**

So, the overall spatial variance at the detector would be greater than that at the barrel, $\langle {}^2 x_D \rangle > \langle {}^2 x_B \rangle$; and, the change in shape of the spectrum at the detector would mean that a new spatial correlation by energy had appeared.  The total energy (neutrino flux) delivered to the target would not change as a function of neutrino mass, because we are counting only good detector events:  Neutrinos are not known to decay, so every neutrino detected will have left the storage ring with the energy at which it was counted.

We may define the shape of the neutrino event-count spectrum $\Psi_R(x, E)$ at a point $x_R$ in the barrel of the storage ring by combining the expected energy distribution with the luminosity decay function for muons as,

$$\Psi_R(x_R, E) = \begin{cases} \mathbf{G}(E) \exp\left( \dfrac{-x_R}{t_m c \sqrt{g_m^2 - 1}} \right), & t_{x_R} \subset T_R \\ \\ 0, & \text{otherwise} \end{cases}, \tag{8}$$

within some storage-ring phase factor for $T_R$ (see Table 1), and in which the *m* subscripts identify muon-specific parameters.

Using Eq. (8), we have plotted the upper half of Fig. 2 for $x_R \cong 0$ again in Fig. 3 in a somewhat different way.   As shown in Fig. 3, the 250-m long muon train has just finished entering the barrel and so the neutrino creation event count has gone to 0 at $x_R$.   If we chose a storage ring phase so that $f = 0$ marked the arrival of the first muons, the time shown would be about $f = p/4$, the $t \subset T_R$ regime over the barrel being shown about $1/4$ elapsed according to the Table 1, row 3 schedule.

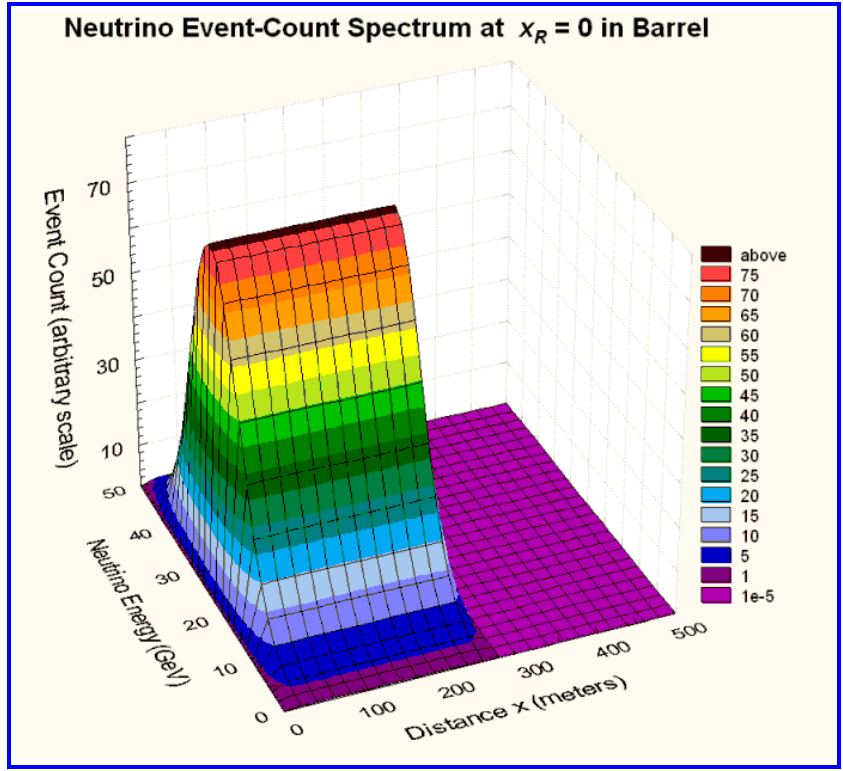

**Figure 3.  The spectrum $Y_R(x, E)$ past the point of creation $x_R$ of a neutrino in the 50-GeV muon storage ring barrel, at phase $f = p/4$ during regime $t \subset T_R$.  Compare with the upper half of Fig. 2.  The event count is scaled arbitrarily for readability.**

Turning to the image spectrum at the detector, for a neutrino creation event known to have occurred exactly at $x_R$ in the barrel, consider the corresponding image event located at $x_D$.   On the light cone, we have the time of flight,

$$t_c = \frac{(x_D - x_R)}{c} = \frac{\Delta x}{c}.$$  (9)

Assuming a massive particle moving at speed $u$ in the lab frame, we may use the relativistic formula, $E/c = m\sqrt{(gu)^2 + c^2}$, to solve for $u$ and compute the lab frame time of flight $t_{D-R}$ as,

$$t_{D-R} = \frac{\Delta x}{u} = \frac{\Delta x}{c} \frac{E}{\sqrt{E^2 - (mc^2)^2}} \; ; \tag{10}$$

so, at the detector we must resolve a lab-frame time-difference $\Delta t$ such that,

$$\Delta t = \Delta T_R(E) = t_{D-R} - t_c = \frac{\Delta x}{c}\left(\frac{E}{\sqrt{E^2 - (mc^2)^2}} - 1\right). \tag{11}$$

Thus, the lab-frame flight time of a neutrino of rest mass $m$ and energy $E$ must differ from the corresponding light-cone time on the interval $\Delta x$ by $\Delta T_R(E)$. So, the lab-frame displacement $d\!k$, of a neutrino of energy $E$ from the light-cone image of its creation point, given a fixed distance $\Delta x$ between barrel and detector, may be written as,

$$d\!k(E) = u \Delta T_R(E) = u \frac{\Delta x}{c}\left(\frac{E}{\sqrt{E^2 - (mc^2)^2}} - 1\right) = \Delta x\left(1 - \sqrt{1 - \frac{(mc^2)^2}{E^2}}\right), \tag{12}$$

in which we have substituted $\sqrt{1 - (mc^2/E)^2}$ for $u/c$.

From (12), for each good-event neutrino of energy $E$, the image will have been shifted by $d\!k(E)$ relative to where it would be on the light cone. Corresponding to a barrel point $x_R$, the set $\{x_D\}$ of all corresponding image events of energy $\{E\}$, translated back on the light cone to the origin on the barrel, will be given by

$$\{x_D(x_R)\} = x_R - d\!k(\{E\}) = x_R - \Delta x\left(1 - \sqrt{1 - \frac{(mc^2)^2}{\{E^2\}}}\right), \tag{13}$$

noticing that for massless neutrinos (13) reduces to $x_D = x_R$, as it should.

Eq. (13) describes the mapping $x_R \xleftrightarrow{\Delta x} x_D$ described *ca.* Fig. 2. For any specific $x_D \equiv \hat{x}_D$,

$$\hat{x}_D = x_R - \Delta x \left( 1 - \sqrt{1 - \frac{(mc^2)^2}{E^2}} \right). \tag{14}$$

As we vary $x_R$ in (14), the value of $E$ also must be changed to keep $x_D$ constant; however, any $\hat{E}$ in the spectrum yields a valid $\hat{x}_D$ for every valid $x_R$. Also, $E$ remains the same between every image point $x_D$ and its barrel point $x_R$. So, for any point in the barrel, to obtain the image spectrum $\Psi_D$ from the barrel spectrum $\Psi_R$, we need only use (14) to substitute for $x$ in an expression for $\Psi_R(x, E)$.

Although the shape of the spectrum in the barrel is very close to constant, its intensity changes with the phase of the storage ring. From here on, when referring to the geometry of the storage-ring barrel itself, the neutrino event-count spectrum will be denoted $\Psi_R$; when referring to the barrel spectrum to be imaged at the detector, $\Psi_B$ will be used in order to include also the entry and exit extent of the 250 m muon train. $\{\Psi_R\}$ covers 500 m; $\{\Psi_B\}$ covers 750 m.

As *ca.* (5) and (6) above, the barrel and its image may be treated in three segments, rising, uniform, and falling, each including a correction for muon decay in the barrel. In the barrel, or on the light cone to the detector, we have the event count spectrum,

$$\Psi_B(x, E) = \Psi_R(\{x_{\text{rise}}\}, E) + \Psi_R(\{x_{\text{uni.}}\}, E) + \Psi_R(\{x_{\text{fall}}\}, E); \tag{15}$$

or, using (8) to define the entire $L_n(x)$,

$$\Psi_B(x, E) = \frac{\mathcal{G}(E)}{250} \exp\left( \frac{-x}{t_m \cdot c \cdot \sqrt{g_m^2 - 1}} \right) \cdot \begin{cases} x, & 0 \leq x < 250 \\ 250, & 250 \leq x < 500 \\ 750 - x, & 500 \leq x \leq 750 \end{cases}. \tag{16}$$

To standardize (16) to a unit of energy, we define $\Sigma_B(x, E) = \frac{1}{K_E} \Psi_B(x, E)$, using

$$K_E = \int_1^{50 \text{ GeV}} dE \int_0^{750 \text{ m}} dx \cdot \Psi_B(x, E) \cong 498.099. \tag{17}$$

The resulting barrel spectrum is shown in Fig. 4.

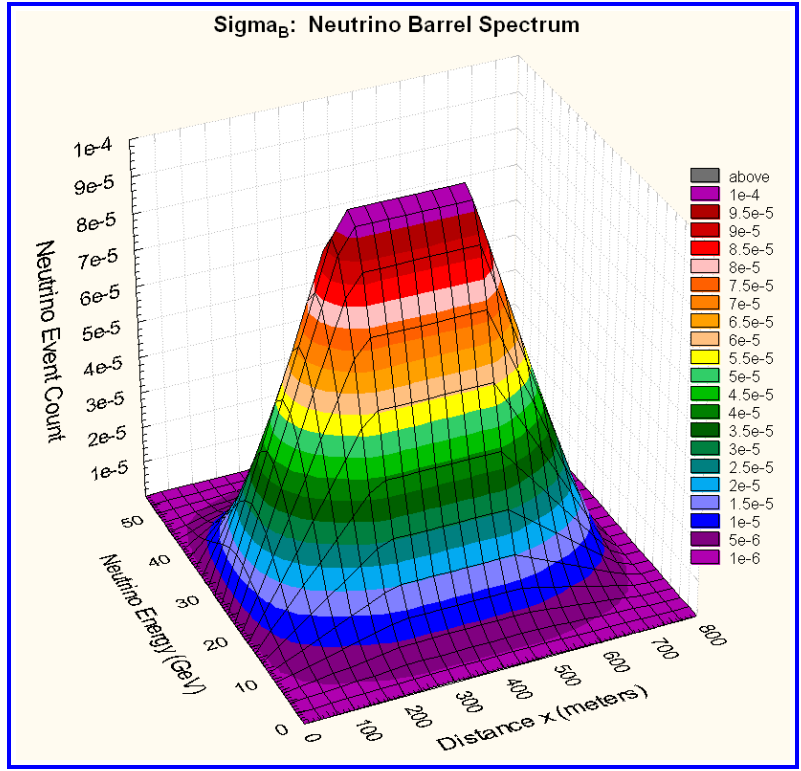

**Figure 4. The standardized barrel spectrum $\Sigma_B(x, E)$ of a neutrino from a 50 GeV muon storage ring 1.5 km in circumference and operated at 50% duty cycle. The volume is standardized to unity. A short-baseline detector is assumed. The image spectrum at the detector, $\Sigma_D(x - \Delta x, E)$, would be indistinguishable from $\Sigma_B(x, E)$ at the scale of this figure.**

At the detector, we may compute the image spectrum by substituting (14) into (16), except in the muon decay factor, to obtain,

$$\Psi_D(x, E) = \frac{\mathcal{G}(E)}{250} \exp\left(\frac{-x}{t_m \cdot c \cdot \sqrt{g_m^2 - 1}}\right) \cdot \begin{cases} x - \Delta x\left(1 - \sqrt{1 - \frac{(mc^2)^2}{E^2}}\right), & 0 \le x < 250 \\ 250 - \Delta x\left(1 - \sqrt{1 - \frac{(mc^2)^2}{E^2}}\right), & 250 \le x < 500 \\ 750 - x - \Delta x\left(1 - \sqrt{1 - \frac{(mc^2)^2}{E^2}}\right), & 500 \le x \le 750 \end{cases} \quad (18)$$

Energy being conserved, we may standardize (18) to $\Sigma_D$ using the same $K_E$ as for $\Sigma_B$ above.

As a minor detail, recall that the barrel spectrum interval $\{x_B\}$ is fixed by the $t \subset T_R$ regime at 750 m long; it maps to the detector as an image approximately on $\Delta x < x_D < \Delta x + 750 \text{ m}$. Neutrinos can not propagate faster than light, so the image spectrum is bounded above by something less than $\Delta x + 750 \text{ m}$, and below by about the same something less than $\Delta x$. The neutrino speed will be much greater than that of the 50 GeV muons, guaranteeing the neutrino $g \gg 500$. The lab-frame difference at either end of the image at $\Delta x = 5 \text{ km}$ would be about 1 cm at $g = 500$; so, we shall ignore all this in our calculations and treat the barrel and image domains as equal.

We wish to calculate the difference spectrum, $\Sigma_\Delta(x, E) \equiv \Sigma_D(x, E) - \Sigma_B(x, E)$, by which we hope to estimate the neutrino mass. Inspection of (18) suggests the simplification,

$$\Psi_D(x, E) = \Psi_B(x, E) - \frac{\mathcal{G}(E)}{250} \exp\left(\frac{-x}{t_m \cdot c \cdot \sqrt{g_m^2 - 1}}\right) \cdot F(E). \tag{19}$$

in which we have written $F(E) \equiv F(E; \Delta x, m)$ in place of the imaging factor in (18). The main reason for factoring out $F(E)$ is that, for low-precision PC software, it may be evaluated separately by a Taylor expansion around the small rest energy factor.

We immediately see that the difference spectrum we want will be given by,

$$\Sigma_\Delta(x, E) = \frac{-1}{K_E} \frac{\mathcal{G}(E)}{250} \exp\left(\frac{-x}{t_m \cdot c \cdot \sqrt{g_m^2 - 1}}\right) \cdot F(E), \quad 0 \leq x \leq 750 \text{ m}, \tag{20}$$

Because we implicitly standardized $\Psi_B(x, E)$ and $\Psi_D(x, E)$ in (19) to obtain (20), $\Sigma_\Delta(x, E)$ of (20) must be the desired, standardized result.

The result is plotted in Fig. 5. This local structure was somewhat unexpected. The greater slope at the lowest energies indicates the direction in which technology must move to employ a nonoscillating, accelerator-based measurement of the neutrino mass. The astronomical amplification factor of $10^{21}$ suggests a statistical test will be hopeless, but we try anyway.

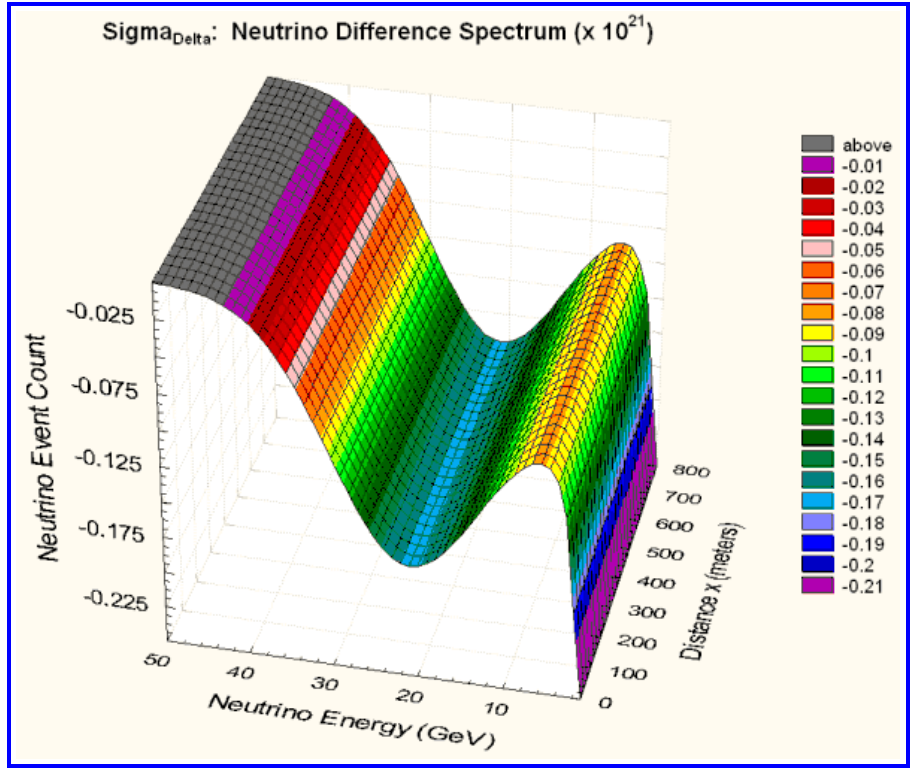

**Figure 5. The standardized difference spectrum $S_D(x, E)$ ´ $10^{21}$ of a $10\,eV/c^2$ neutrino as a function of neutrino energy and location in the barrel image. The event count is from the standardized spectra as in Fig. 4; so, it is a difference between two spectra each ~ $5 \cdot 10^{-5}$ on the unscaled ordinate. The neutrino propagation distance was 5 km.**

Clearly, $\Sigma_\Delta(x, E) \to 0$ in (20) for massless neutrinos, and the calculated effect in Fig. 5 obviously is not zero. The question is, how big a statistical sample of good neutrino events would we need to use it to prove the mass of a 10 $eV/c^2$ neutrino?

Let us assume we will be satisfied to detect mass while risking an error at statistical significance level $p < .001$. Naively, we shall attempt to answer the question of mass by a chi-square ($c^2$) analysis.

Accepting a 20% energy accuracy at the detector, we begin by partitioning the energy domain into mutually exclusive categories: Starting at the high end of 50 GeV, we use $E_i = 4E_{i-1}/5$ to define 18 energy categories. This is the best resolution we can expect in the direction of the structure in Fig. 5; so, we shall pool all distances $x$ on the image to reduce variance within each category.

There is no free parameter to fit the data. Therefore, from a table, $c^2_{.001}(17) = 40.8$. So, we wish to compare the value 40.8 with,

$$c_{\text{data}}^2(17) \equiv \sum_{i \in [1,18]} \frac{\Delta_i^2}{s_i^2} = \sum_{i \in [1,18]} \frac{\Delta_i^2}{s_i^2/N_i}, \tag{21}$$

which is just the sum of squares of 18 unit normal deviates, each with mean $m_{D-B} \equiv 0$ and variance $s_{D-B}^2$ as given by the null-hypothesis difference spectrum, $\Sigma_D - \Sigma_B$ ca. Eq. (19).  The $\Delta_i$ in (21) represent the difference-spectrum $\Sigma_\Delta(x, E)$ of (20) at the category means $E = \langle E \rangle_i$.  We are assured by the central limit theorem that the category samples will be normally distributed with standard deviation $\sqrt{s_i^2/N_i}$, in which $N_i$ is the number of data in the $i$-th category.

The muon-decay dependence on distance approximately has vanished in Fig. 5, and it certainly is zero under the null hypothesis, so we shall ignore it.

To treat the categories accurately according to the expected neutrino energy distribution, we find the weighted average midpoint of each category.  We also find the variance in each category to standardize the data in it to unit normal.  For the variance $s^2$, we directly integrate $\mathcal{G}(E)$ appropriately.

We then compute the category means $\langle E \rangle_i$ of the difference spectrum $\Sigma_\Delta(E)$.  The result is in Table 2.

**Table 2. A chi-square test for the mass of a 10 $eV/c^2$ neutrino from a 50 GeV muon storage ring 5 km distant, assuming a sample of $10^{10}$ good neutrino events. The test is against the difference spectrum $\Sigma_\Delta$ of Eq. (20). Values are scaled and rounded for readability.**

| $\langle E \rangle$ | Flux $N$ ($\times 10^{-5}$) | $H_0 : s_{\{E\}}$ | $H_0 : \dfrac{s_{\{E\}}}{\sqrt{N}}$ ($\times 10^5$) | $m_{\Sigma_\Delta} = \Sigma_\Delta(\langle E \rangle)$ ($\times 10^{21}$) | $c^2_{\text{data}}$ Element $= N\left(m^2_{\Sigma_\Delta} / s^2_{\{E\}}\right)$ ($\times 10^{37}$) |
|---|---|---|---|---|---|
| 43.0 | 5019.03 | 2.392 | 10.68 | -0.007 | 0.04 |
| 35.3 | 21445.59 | 2.206 | 4.76 | -0.047 | 9.74 |
| 28.7 | 30425.74 | 1.826 | 3.31 | -0.119 | 129.16 |
| 23.2 | 22357.88 | 1.464 | 3.10 | -0.166 | 287.31 |
| 18.6 | 11557.91 | 1.169 | 3.44 | -0.167 | 236.03 |
| 14.9 | 5115.70 | 0.934 | 4.13 | -0.144 | 121.55 |
| 11.9 | 2178.94 | 0.748 | 5.07 | -0.118 | 54.20 |
| 9.5 | 952.21 | 0.600 | 6.15 | -0.101 | 27.00 |
| 7.6 | 440.52 | 0.481 | 7.24 | -0.092 | 16.13 |
| 6.1 | 218.33 | 0.385 | 8.25 | -0.089 | 11.64 |
| 4.9 | 116.05 | 0.309 | 9.06 | -0.097 | 11.45 |
| 3.9 | 65.85 | 0.247 | 9.64 | -0.100 | 10.76 |
| 3.1 | 39.61 | 0.198 | 9.95 | -0.121 | 14.79 |
| 2.5 | 25.05 | 0.159 | 10.01 | -0.147 | 21.55 |
| 2.0 | 16.54 | 0.127 | 9.86 | -0.189 | 36.71 |
| 1.6 | 11.32 | 0.102 | 9.54 | -0.259 | 73.67 |
| 1.3 | 7.98 | 0.081 | 9.10 | -0.326 | 128.47 |
| 1.0 | 5.76 | 0.065 | 8.56 | -0.509 | 353.25 |

$$c^2_{\text{data}} = \sum_{\langle E \rangle} c^2_{\text{data}} \cong 1.5 \cdot 10^{-34}$$

$$c^2_{.001} \cong 40.8$$

### Conclusion

This paradigm appears inapplicable to high-energy neutrinos. Other tests based on, say, cross-correlation might use the available variance better than chi-square, but it appears that they would not come close to revealing neutrino mass under the differential time paradigm. The spectral shape difference at 5 km would seem to be too minuscule to be useful.

## *Absolute Time Paradigm*

In this case, laser calibration or other computation of the speed of light would be used to establish the distance along the light cone from the barrel to the detector. The arrival or exit time in the barrel of a train of muons then would mark the start time of an absolute time-of flight measurement of the speed of the neutrinos. If a sample mean of the corresponding detector events should be shown to have arrived at the detector at the speed of light, the neutrinos would be shown massless.

We put aside for now the obvious problem of determining the locus of creation of an individual neutrino and merely calculate some rough limits on the usefulness of this paradigm:

The calculations already have been done above, and the relevant result is in Eq. (9): Expanding around the small rest-energy term, we may plot the required resolution for a rather heavy $10\,\text{eV}/c^2$ neutrino:

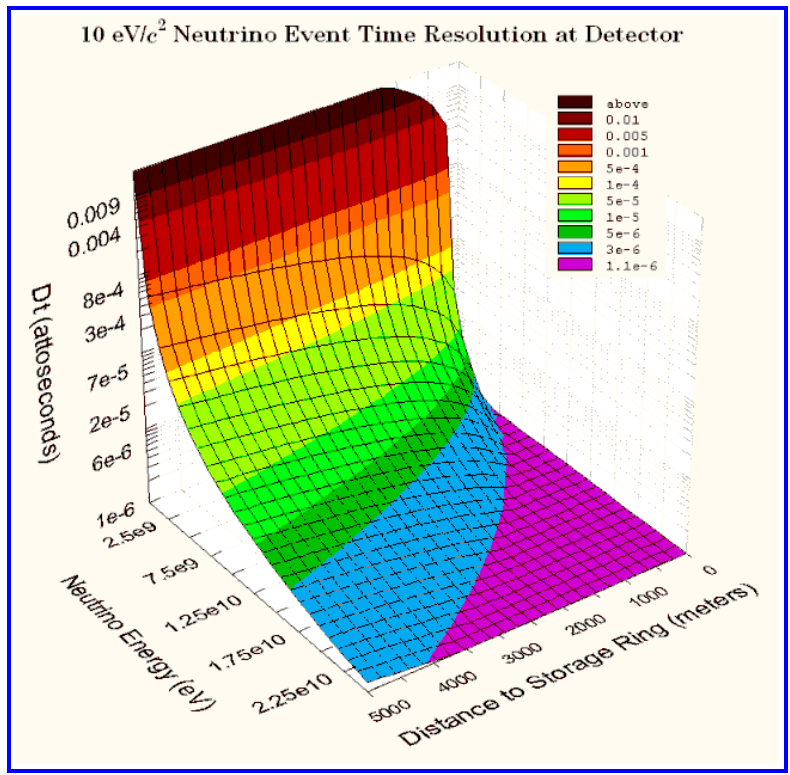

**Figure 6. The point-event detector time resolution required in Eq. (9) to estimate the mass of a $10\,\text{eV}/c^2$ neutrino. Note that the vertical scale is in attoseconds ($10^{-18}$ s).**

The neutrino problem seems hopeless in this paradigm, too. The uncertainty principle would begin to dominate experimental error at the shorter distances and the higher energies shown in Fig. 6, even were there no random or systematic error at all.

From Fig. 6, picosecond time resolution at the detector would not save the day even at $\Delta x = 5{,}000$ km, because a 10 $\text{eV}/c^2$ neutrino still would require a million years at $10^{10}$ good events per year. It would take a Mighty Murine, indeed, to apply this paradigm at a practical distance such as that of the planet Pluto!

A fantastic barrel-plus-detector system capable of resolving a time difference down to a single wavelength of visible light at, say, 600 nanometers, would correspond to a time resolution of $600 \cdot 10^{-9}/3 \cdot 10^8 = 2$ fs $= 2 \cdot 10^{-3}$ ps. In Fig. 6, at 5 km, for a rather heavy 10 eV/$c^2$ neutrino, this would imply a sample size of about $\sqrt{N} = 10^8$, or, a tally of some $10^{16}$ good events. Another million-year study, even assuming that such a time-resolution was feasible.

However, a similar experiment with one-wavelength resolution carried out at Earth-Moon distance would reduce the required event count by $\sqrt{N} \sim 10^5$. We no longer would have a short-baseline experiment, but this setup would seem to render credible a meaningful improvement on the current upper bound of the neutrino mass.

Looking at a more massive particle for comparison, Fig. 7 shows that the same Fig. 6 calculation for a particle as heavy as a muon, at a line-of-sight range up to 5 km and at nanosecond time resolution, would require only a handful of events to prove it massive. Even at rather coarse *ns* time resolution, squeezing off a sample of a few times $10^6$ good events would be enough to estimate this mass to within a factor of two.

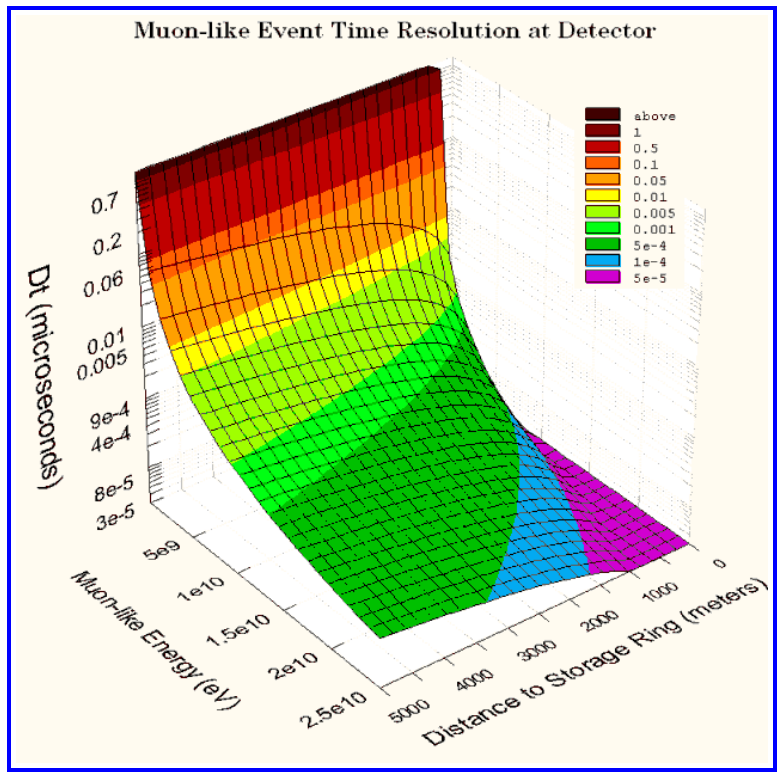

**Figure 7. The point-event detector time resolution required in Eq. (9) to estimate the mass of a particle around the 106 MeV/$c^2$ mass of a muon. The time scale is in microseconds.**

Conclusion

Counter-intuitively, the absolute-time paradigm comes closer to feasibility than the differential-time paradigm. Seemingly, though, to make the absolute-time paradigm practical for ~25 GeV neutrinos, one would have to improve the precision of measurement on a short-baseline interval by some 9 to 12 orders of magnitude beyond that reasonably attainable.

An Earth-Moon baseline would seem to be in the realm of the possible; radiation hazards of course would be avoided entirely if the storage ring were located on the Moon.

The sensitivity of the absolute-time paradigm depends on neutrino speed or time-of-flight measurement. Because a change in energy is equivalent to a squared change in speed, 12 orders of magnitude of speed improvement at 25 GeV would imply 0.025 MeV neutrinos. It is unclear how an equivalent beam of ~$10^{10}$ good events per year could be accomplished for 25 KeV neutrinos at a distance of ~5 km. It would be interesting to know whether the presently proposed muon storage ring, or apparatus associated with it, could be designed or operated for these energies.

# Acknowledgements

This paper was composed in *Microsoft Word*. Drawing was done in *CorelDRAW*, and graphs were plotted in *Statistica*. Some calculus was verified in *MathCAD*, which also was used for numerical integration and series expansion. Some high-precision arithmetic was verified with the programmable `bc` application, *PAPC for Windows*. The completed manuscript was printed to a postscript file from which PDF format was written using *Adobe Acrobat*. Italicized words are names used in trade by the respective product owners.